\documentclass[reprint,aps,prl,superscriptaddress,floatfix,nobibnotes,nofootinbib]{revtex4-2}

\usepackage{amsmath,amssymb,bm}
\usepackage{graphicx}
\usepackage{xcolor}
\usepackage{upgreek}
\usepackage{siunitx}

\usepackage{algorithm}
\usepackage{algpseudocode}

\usepackage[caption=false]{subfig}

\usepackage{xspace}
\usepackage[colorlinks=true,allcolors=blue]{hyperref} 
\usepackage{placeins} 

\usepackage{booktabs}

\newcommand{\mpbax}{multipointBAX\xspace}

\begin{document}

\title{Efficient Dynamic and Momentum Aperture Optimization for Lattice Design Using Multipoint Bayesian Algorithm Execution}

\author{Z.~Zhang}
\affiliation{SLAC National Accelerator Laboratory, Menlo Park, CA 94025}
\author{I.~Agapov}
\affiliation{Deutsches Elektronen-Synchrotron DESY, Hamburg, 22607, Germany}
\author{S.~Gasiorowski}
\affiliation{SLAC National Accelerator Laboratory, Menlo Park, CA 94025}
\author{T.~Hellert}
\affiliation{Lawrence Berkeley National Laboratory, Berkeley, CA 94720}
\author{W.~Neiswanger}
\affiliation{University of Southern California, Los Angeles, CA 90089}
\author{X.~Huang}
\affiliation{SLAC National Accelerator Laboratory, Menlo Park, CA 94025}
\author{D.~Ratner}
\email{dratner@slac.stanford.edu}
\affiliation{SLAC National Accelerator Laboratory, Menlo Park, CA 94025}

\date{\today}

\begin{abstract}

We demonstrate that multipoint Bayesian algorithm execution can overcome fundamental computational challenges in storage ring design optimization. Dynamic (DA) and momentum (MA) optimization is a multipoint, multiobjective design task for storage rings, ultimately informing the flux of x-ray sources and luminosity of colliders. Current state-of-art black-box optimization methods require extensive particle-tracking simulations for each trial configuration; the high computational cost restricts the extent of the search to $\sim$$10^3$ configurations, and therefore limits the quality of the final design. We remove this bottleneck using \mpbax, which selects, simulates, and models each trial configuration at the single particle level. We demonstrate our approach on a novel design for a fourth-generation light source, with neural-network powered \mpbax achieving equivalent Pareto front results using more than two orders of magnitude fewer tracking computations compared to genetic algorithms. The significant reduction in cost positions \mpbax as a promising alternative to black-box optimization, and we anticipate \mpbax will be instrumental in the design of future light sources, colliders, and large-scale scientific facilities.

\end{abstract}

\maketitle

Designing modern scientific facilities — from synchrotron light sources to particle colliders — requires optimizing hundreds of parameters in a complex, nonlinear systems, where a single design evaluation can take hours of computation. In storage rings, this challenge is exemplified by dynamic aperture (DA) and momentum aperture (MA) optimization, where maximizing the regions of particle stability directly determines injection efficiency, beam lifetime, and ultimately the photon flux or luminosity achievable in next-generation facilities.

The computational bottleneck is severe: maximizing DA and MA is a type of multipoint optimization, where evaluating a single lattice design requires tracking tens of thousands of particles for hundreds of thousands of turns, making global optimization prohibitively expensive.  Moreover, there is a trade-off between maximizing DA and MA area, so the standard approach is to find a Pareto front; i.e. DAMA optimization is not just multipoint, but also multiobjective. Current state of the art methods, e.g. multiobjective genetic algorithms (MOGA) and multiobjective particle swarm optimization (MOPSO) \cite{borland2009direct,YANG200950,HUANG201448}, typically explore $\sim 10^3$ designs, forcing compromises between optimization scope and computational cost that result in suboptimal performance — a limitation that has been identified as a grand challenge for the field \cite{osti_1764152, nagaitsev2022accelerator}. 

We present a paradigm shift by combining multipoint Bayesian Algorithm Execution (\mpbax) \cite{miskovich2024multipoint} with deep neural network surrogates to achieve two-orders-of-magnitude speedup over state-of-the-art (SOTA) methods. Our approach enables simultaneous optimization of dozens of lattice parameters while tracking particles in up to millions of potential configurations, which previously was computationally intractable.

Applied to the SSRL-X storage ring lattice, our method discovers Pareto-optimal solutions with $<$1\% of the function evaluations needed by SOTA methods. This demonstration highlights how modern machine learning can overcome fundamental computational barriers in large-scale scientific facility design, with immediate implications for current facility upgrades and the next generation of light sources and colliders.

\textit{Background and related work:} In classical DAMA optimization, for each proposed design configuration, thousands of particles with different initial transverse positions (for DA) and momenta and longitudinal positions (for MA) are simulated for thousands to hundreds of thousands of turns around the storage ring.
For DA, particles are typically chosen along rays of increasing distance from the origin, and the particle with maximum radius that survives the full number of turns defines the outer edge of the DA. The metric is the area enclosed by those points. At the end of the optimization, each potential solution is then re-evaluated using a random collection of `error seeds,' i.e. random changes to the design configuration that approximate a realistic ring's performance.  A similar process is applied to the longitudinal positions and momentum offset for MA \cite{steier2002measuring,kim2022hybrid}. Ultimately, our goal is to identify a Pareto front of solutions in the two-dimensional DA and MA space.

The classical SOTA approaches for DAMA optimization, e.g. MOGA\cite{murata1995moga, yang2009global} and MOPSO\cite{coello2002mopso}, are slow for two reasons: first, they require many iterations to converge; and second, each iteration involves simulating thousands of particles to calculate the objectives. These are sometimes referred to as `outer' and `inner' loop problems respectively \cite{stevens2020ai}.  ML methods have been explored as solutions to both challenges.

Outer-loop approaches reduce the number of iterations by learning a surrogate model of the objective. For example, Bayesian optimization (BO) uses a surrogate to construct an acquisition function to guide sampling. BO has been widely adopted for accelerator applications \cite{roussel2024bayesian}, including for DAMA optimization \cite{huang2021multi}. Similarly, \cite{li2018genetic} uses K-means clustering (rather than a Gaussian process and BO) to improve population fitness within MOGA. 
However, there is a ceiling on the efficiency gains of simply reducing the number of iterations --- even a small number of iterations is costly due to the expense of each inner-loop acquisition.

Inner-loop approaches use fast ML surrogates to replace some or all of the optimization procedure's calls to the expensive physics simulation, reducing the average cost of each iteration. Because surrogates can be orders of magnitude faster than physics simulations, calls to the surrogate are effectively free. For example, even using an inefficient optimizer (e.g. MOGA), computational expense is reduced by replacing the physics simulator with a neural network (NN). These surrogates can either model the DAMA objective, i.e. directly predict the survival area of the DA and MA maps \cite{kranjvcevic2021multiobjective, wan2020neural, lu2021enhancing, lu2023demonstration}, or can model the full maps, i.e. predict the survival time for each individual particle simulation \cite{schenk2021modeling, van2021using, wan2022machine, di2024accelerating}. (Relatedly, NNs can also extend DAMA estimates to large numbers of turns beyond the limit of conventional tracking \cite{van2021using, casanova2023ensemble, wan2022machine} and identify outliers \cite{van2021using}.) 

For NN approaches, the assembly of the training set often dominates the computational cost. Training on static, pre-existing datasets risks that the surrogate will be inaccurate in the optimal region.  Consequently, recent activity has focused on active learning, i.e., continually acquiring training data during the training and/or optimization processes \cite{wan2020neural, huang2021multi, kranjvcevic2021multiobjective, lu2023demonstration, di2024optimizing}, typically selecting an entire map at each iteration. Still, after amortizing the cost of the training dataset, the overall speedup compared to SOTA methods is typically modest.

This paper presents a new path to DAMA optimization, using the outer loop to maximize efficiency of the inner-loop training dataset. The work involves the following innovations: (1) We select simulations and train the surrogate at the granularity of single particles, rather than using full maps.
(2) We use an algorithmically-driven acquisition function (i.e. \mpbax) which selects individual points that maximally impact the optimization outcome, rather than just reducing model uncertainty. 
(3) We implement a NN-based BAX framework, avoiding the $n^3$ scaling of the Gaussian processes used by previous BO and BAX applications in accelerator physics. The NN approach is critical to handle the $>$100k points modeled in DAMA optimization.

\textit{Approach} Multipoint optimization refers to tasks in which each iteration of the optimizer requires a scan of variables in a secondary space. The term originated in aeronautic design (e.g. see \cite{nemec2004multipoint, kenway2014multipoint, liem2015, CHAI201899}), but is common in scientific fields in general (e.g. see \cite{yamashita, terayama, miskovich2022recoil}), with multiple examples in recent years for accelerators \cite{miskovich2024multipoint, roussel2024autonomous, matzoukas25efficient}. In DAMA optimization, each acquisition in the configuration space requires a secondary scan of particles in position and momentum space. The expense of the secondary scan --- in DAMA each acquisition requires thousands of particle tracking simulations to return a scalar objective --- is the crux of why traditional multipoint optimization is difficult.

\begin{figure}[htbp]
    \centering
    \includegraphics[width=\linewidth]{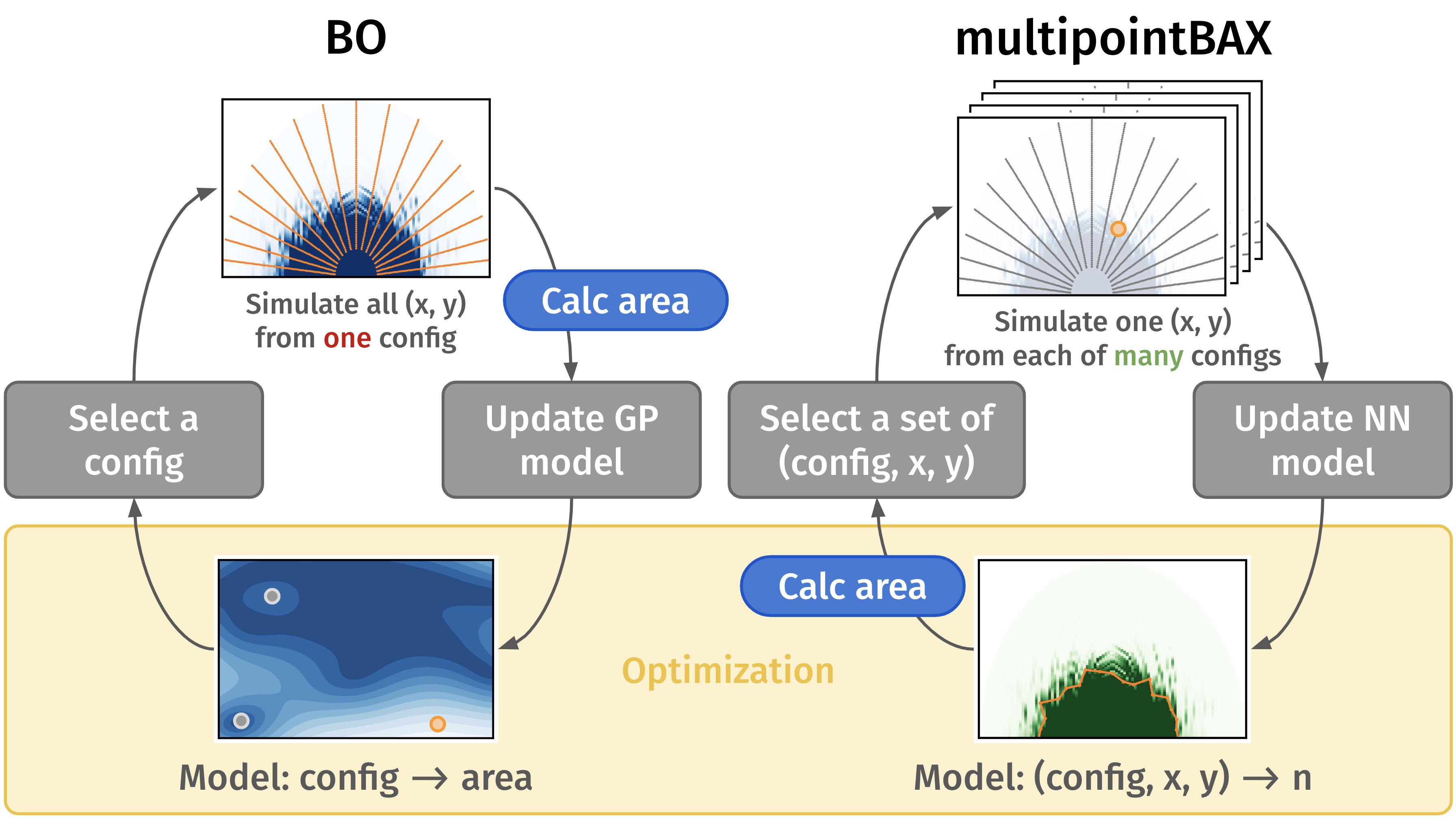}
    \caption{Comparison of Bayesian optimization and BAX for a single objective optimization. There are two key differences: first, in \mpbax we simulate a single particle at each acquisition, whereas BO simulates a full map of more than one thousand particles. Second, the BO surrogate models the objective function (configuration $\rightarrow$ objective), whereas the BAX surrogate models the underlying physics function (i.e. configuration $\rightarrow$ map). Because of the increased complexity and large number of data points to model, BAX also requires use of a NN model whereas BO can use a simpler Gaussian process model.}
    \label{fig:bax_vs_bo_schem}
\end{figure}

To address the multipoint challenge, 
\mpbax trains a surrogate model to construct the full DA and MA stability maps, i.e. the surrogate model will predict the number of turns a particle will survive as a function of both configuration (magnet values) and position (x-y coordinates for DA and s-$\delta$ coordinates for MA). Scanning over the position coordinates then constructs the maps, from which we can calculate the area metrics analytically. We can optimize directly on the cheap surrogate model, calculating the DA and MA area metrics for each configuration, to predict the Pareto front. This procedure (DAMA calculation plus Pareto front finding) is the ``algorithm'' in \mpbax \cite{neiswanger2021bayesian}. Finally, \mpbax uses the output of the surrogate model optimization to construct an acquisition function, which selects new points to acquire from the physics simulator. Specifically, the acquisition function tries to reduce the surrogate's uncertainty, but only in regions predicted to have optimal performance.

\begin{figure}[htbp]
    \centering
    \includegraphics[width=\linewidth]{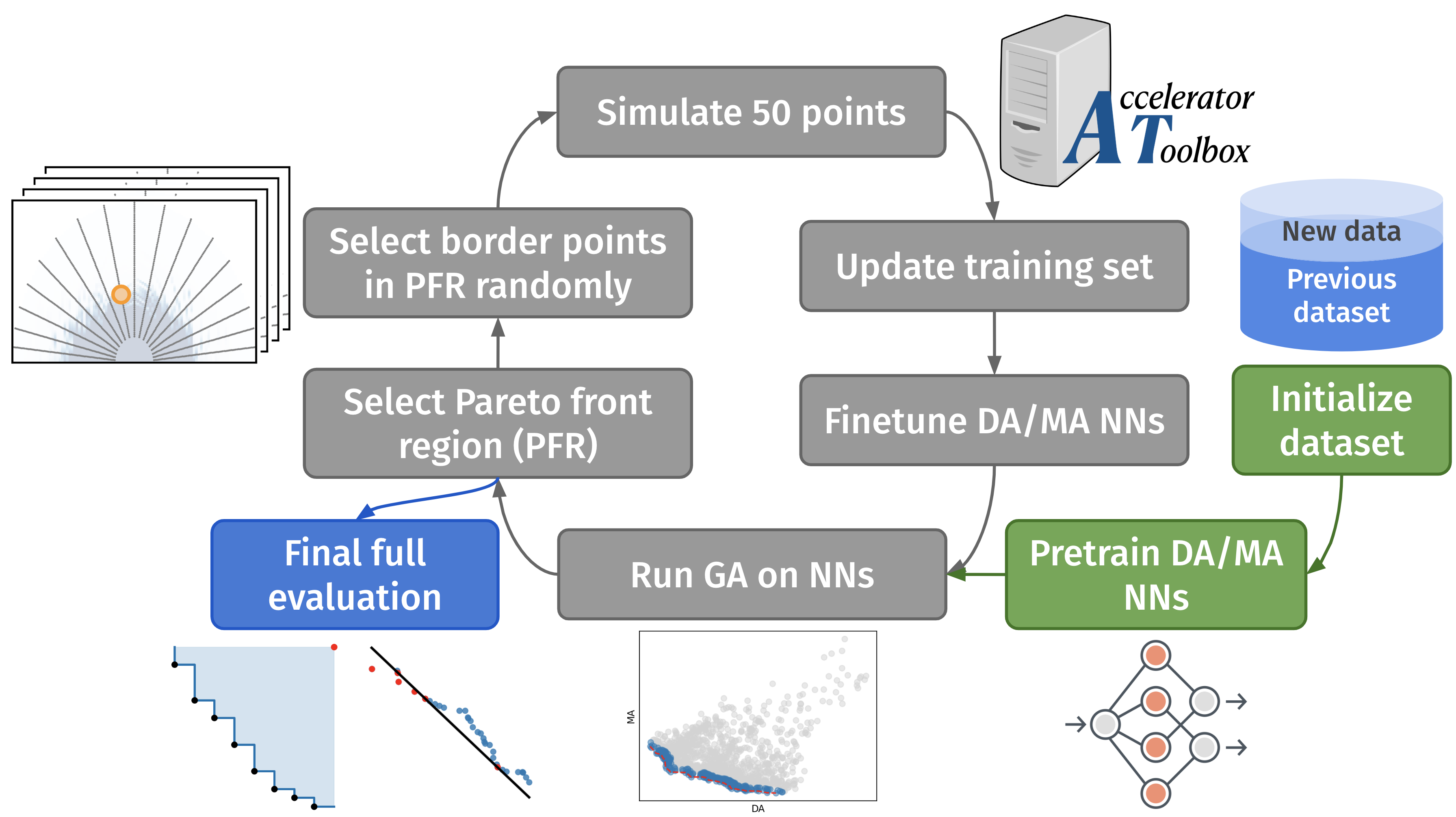}
    \caption{Schematic of the \mpbax implementation for DAMA optimization. For more details, see description in Appendix~\hyperref[appendix:algorithm]{B} and Algo~\ref{alg:mpbax}.}
    \label{fig:full_bax_schem}
\end{figure}

We emphasize that \mpbax never acquires a full DA or MA stability map; each acquisition returns the number of turns for a single particle, and the algorithm typically selects only a single particle simulation for each proposed configuration. 
Only when we have finished the full optimization will we evaluate the full DA and MA maps using the physics simulation. Figure \ref{fig:bax_vs_bo_schem} compares \mpbax and BBO, and Fig.~\ref{fig:full_bax_schem} shows the full \mpbax scheme.

Through the multipoint optimization lens, we can now re-interpret the innovations described previously: selecting individual points -- and selecting only those that impact the optimization outcome -- improve the sampling efficiency, while modeling individual particle outcomes improves information efficiency. Given the large number of simulations in each BO iteration, from an information/sampling perspective, \mpbax can be orders of magnitude more efficient.  

On the other hand, the need to model millions of points (rather than thousands of DA areas) makes it infeasible to use Gaussian processes as the underlying surrogate due to the poor $n^3$ scaling in the number of modeled datapoints. Consequently, we use a neural network (NN) as the surrogate model. We have implemented a neural field~\cite{neural-fields} architecture: the network takes as input both the accelerator configuration and a particle position and predicts the survived turns for that one particle. 
Selecting a single particle at each iteration minimizes the total compute hours used, but given the availability of high-performance computing clusters, as well the overhead of retraining the NN and rerunning the GA search on each iteration, to minimize the wall-clock time we launch a batch of 50 physics simulations in parallel in each iteration.
The detailed algorithm and network, as well as the specific acquisition function, are given in Appendix~\hyperref[appendix:algorithm]{B}.

\textit{SSRL-X problem description}
To demonstrate the method, we apply \mpbax to the design of the proposed SSRL-X ring.  SSRL-X is a low-emittance storage ring lattice that is an option for a future upgrade of the Stanford Synchrotron Radiation Lightsource (SSRL). The design has a natural emittance of 86~pm at 4~GeV~\cite{RAIMONDI2024169137}, based on the hybrid six-bend achromat (H6BA) lattice structure~\cite{RAIMONDIprab2023}. The lattice employs several different combinations of H6BA cells to produce long and short straight sections 
to host insertion devices, super-bends, or injection components, so it lacks the multi-fold periodicity aimed for by many 
other low emittance storage ring lattices. 
With the use of the H6BA lattice structure and by enforcing transparency conditions, the lattice achieves a 
high nonlinear-dynamics performance. However, numerical optimization of the dynamic aperture and momentum aperture is still necessary. 

The MOPSO algorithm has been previously used to optimize the dynamic aperture and the Touschek lifetime with six sextupole families~\cite{RAIMONDI2024169137}. In~\cite{RAIMONDI2024169137}, the strengths of the sextupole families are varied freely within the 6-dimensional space such that optimization can change -- and optimize -- the chromaticities~\cite{HuangOnlineDA}. However, additional constraints are necessary to prevent a large positive horizontal chromaticity and a negative vertical chromaticity. 
In the current study, we use the chromaticity response matrix of the 6 sextupole families to create 4 knobs that do not change
chromaticities, in the same manner as found in Ref.~\cite{HuangOnlineDA}. This approach eliminates the need 
to implement additional constraints on chromaticities. Additionally, SSRL-X uses an off-axis injection scheme from the $-x$ side of the horizontal plane. We account for this applying a weight of 2.0 to the $-x$ side of the stable region in the DA objective calculation.

Evaluation of DA and MA in the previous study \cite{RAIMONDI2024169137, RAIMONDIprab2023} is done by tracking thousands of particles launched with initial 
coordinates spanning the phase space (for DA) and from various locations with different energy errors (for MA) for thousands of turns.  The optimization is computationally costly and time consuming,
and typically only the performance for one error seed (for linear optics and coupling errors) is evaluated for each sextupole solution. Using \mpbax can shorten the optimization cycle, allowing more robust optimization across multiple error seeds, and search over larger parameter spaces. 

\begin{figure*}[htbp]
    \centering
    \includegraphics[width=\linewidth]{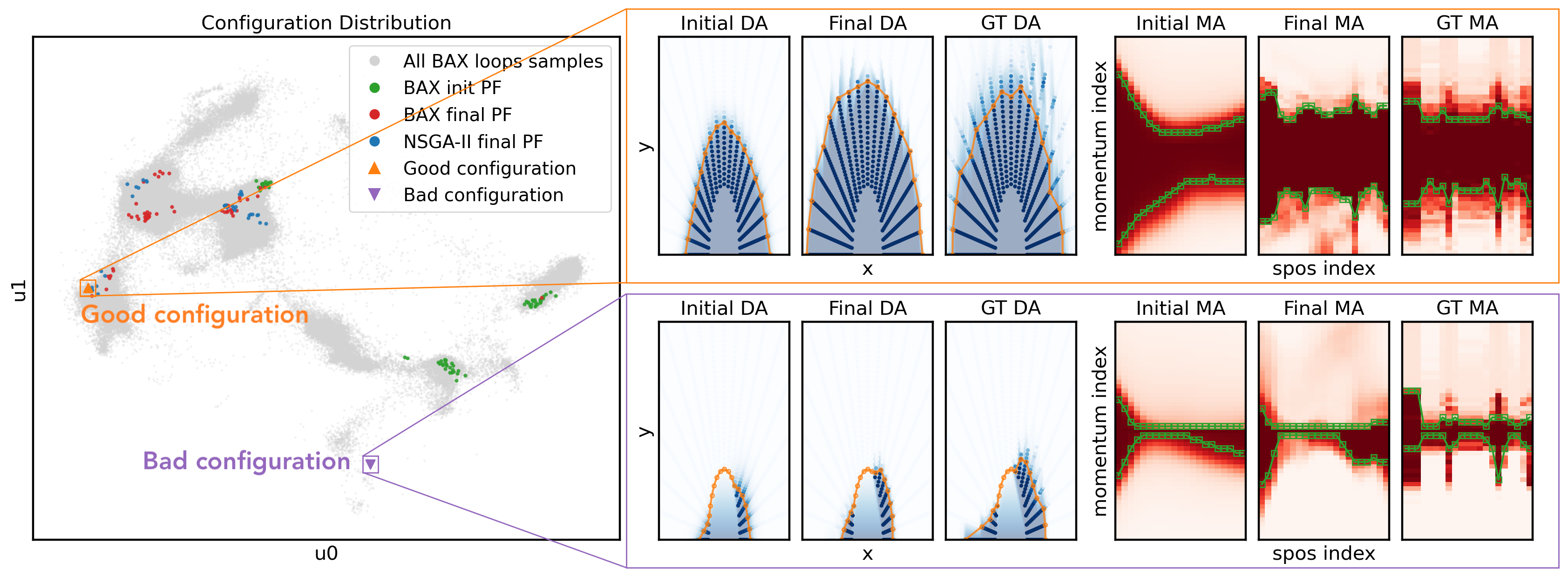}
    \caption{Model learning process: Left scatter plot shows the configuration input distribution projected on two dimensions (u0 and u1) using a UMAP \cite{umap} transform fit on initial samples of BAX. Gray points show all sampled configurations during the BAX run, blue points show final PF from GA, red points show final PF from BAX, and green points show initial PF from BAX. Right heatmaps: Each picture shows DA (left three columns) and MA (right three columns) maps predicted by the NN. `spos' is the position along the accelerator. Top row in the orange box shows a `good' configuration predicted to lie on the Pareto front (orange on the UMAP), while bottom row in the purple box shows a random `bad' configuration (purple on the UMAP) far from the predicted Pareto front. For both DA and MA, from left to right: Prediction after initial training (Initial DA/MA), prediction after final training (Final DA/MA), and groundtruth (GT DA/MA).
    }
    \label{fig:DA_pred_vs_gt}
\end{figure*}

\textit{Comparison of BAX and NSGA-II performance}
The full \mpbax algorithm implemented for this study is shown schematically in Fig.~\ref{fig:full_bax_schem}.
We highlight that \mpbax focuses only on promising regions of parameter space: once the model has determined a region is not promising, \mpbax does not resample those configurations, improving efficiency of training data acquisition. On the other hand, \mpbax continually resamples in regions with high DA and MA, increasing prediction confidence in promising regions (Fig.~\ref{fig:DA_pred_vs_gt}).

\begin{figure}[htbp]
    \centering
    \includegraphics[width=\linewidth]{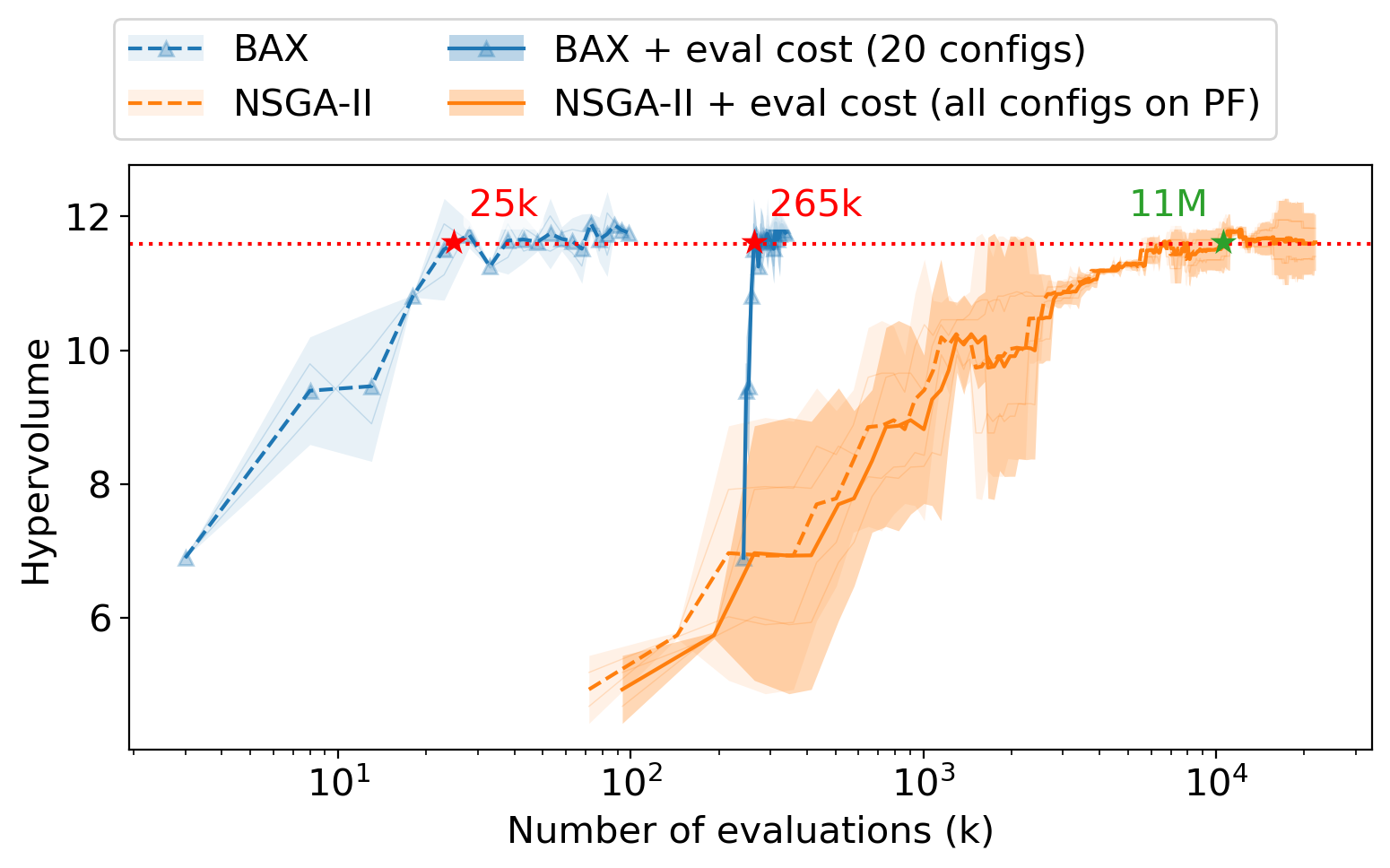}
    \caption{BAX vs. NSGA-II performance as a function of iteration number using hypervolume. We run both BAX and NSGA-II twice, and show the mean performance (dashed and solid curves) with the corresponding $2\sigma$ error bands (light and dim shaded areas). Dashed lines indicate optimization cost, and solid lines include the cost of the final evaluation. Red stars denote BAX’s early-stopping points, and green star indicates where NSGA-II reaches comparable performance.
    }
    \label{fig:BAX_vs_SoTA_HV}
\end{figure}

To assess performance vs. the SOTA, we require both a benchmark algorithm and a metric. We choose NSGA-II (a standard genetic algorithm, with similar performance to MOPSO used in \cite{RAIMONDI2024169137}) as both a SOTA benchmark, and for the internal optimizer within \mpbax. For a metric, we select the hyper-volume (HV) enclosed by the Pareto front and a reference point, with the comparison as a function of simulation queries shown in Fig.~\ref{fig:BAX_vs_SoTA_HV}. 

For both \mpbax and NSGA-II, we randomly select one of 10 error seeds for each call to the physics simulator, which we observe improves the performance of both approaches. After the optimization concludes, we evaluate every point in each final selected configuration using 10 new error seeds, none of which were used during the optimization process. This evaluation cost is included as a separate set of curves in Fig.~\ref{fig:BAX_vs_SoTA_HV}. Note that the evaluation cost for NSGA-II is actually higher than for \mpbax, but the cost is negligible compared to the higher cost of optimization for NSGA-II.

We find \mpbax converges with 40 times fewer acquisitions, all-inclusive, compared to the number of acquisitions required by NSGA-II to reach approximately the same performance (Fig.~\ref{fig:BAX_vs_SoTA_HV}). However, this undersells the efficiency of \mpbax: Only 10\% of the simulations occur during the BAX routine, with the vast majority occurring in the final evaluation after the optimization is complete.
If we only consider the BAX acquisitions, \mpbax converges to an optimum with more than 400 times fewer simulation calls compared to NSGA-II. The high fraction of simulations in the final evaluation is in part due to the relative simplicity of the four-dimensional search. For harder tasks (e.g. higher input dimensionality, more complex behavior), the final evaluation will be a smaller fraction of the computational cost, and the all-inclusive gain will increase. 

While HV is an effective way to compare algorithms, in practice a project must choose a single configuration to build. 
As an additional check, we also compared performance on a single-configuration metric produced by a linear combination of DA and MA. The results, presented in Appendix~\hyperref[appendix:single-metric]{C}, show trends consistent with the HV metric, with \mpbax again yielding all inclusive gains (including final evaluation) that can near a factor of 100.

We implement an early stopping procedure based on convergence of the model's predictions \cite{miskovich2024multipoint}, resulting in the red stars in Fig~\ref{fig:BAX_vs_SoTA_HV}. We also note that Fig.~\ref{fig:BAX_vs_SoTA_HV} compares the computational cost based on the number of simulations acquired, ignoring the shorter run-time of particles that are lost earlier. However, the impact is modest ---NSGA-II simulations are approximately 25\% faster on average --- and does not affect the conclusions. Further details on the early-stopping method and time-cost analysis are provided in Appendix~\hyperref[appendix:early-stopping]{D} and Appendix~\hyperref[appendix:time-cost]{E}, respectively.

Beyond the efficiency gains, \mpbax has an additional advantage of greater interpretability compared to SOTA methods. MOGA and MOPSO provide no information about configurations beyond those sampled during the optimization, and each new query requires an expensive physics simulation. Classical BO algorithms produce a surrogate model of the objective function only. By contrast, the \mpbax routine produces a NN surrogate of the full DA and MA maps, enabling subsequent exploration of new sextupole settings by hand. The NN surrogate provides physicists with intuition about the optimization, for example revealing how the predicted DAMA maps change as a function of sextupole knobs, or suggesting how to promote the growth of DA in desirable areas.

\textit{Conclusion} We have presented an application of \mpbax to DAMA optimization, highlighting the benefit of acquiring and modeling simulations at the individual particle level. We observe 40-90x speed-ups for the proposed SSRL-X lattice while achieving equivalent performance. We anticipate gains could be even higher when applied to more complex optimization tasks, e.g. for the Future Circular Collider \cite{benedikt2020future}, where the optimization is expected to take longer and the final evaluation will be a negligible fraction of compute cost: during the optimization procedure, we observe a factor of more than 400 speed-up compared to NSGA-II.

The implications extend far beyond DAMA. Any multipoint optimization — from beam diagnostics and machine protection at accelerators, to scientific facilities broadly — can benefit from this approach. 
Several advances could amplify these gains further: probabilistic models to quantify uncertainty, improved batch sampling strategies, and pre-training on existing simulation databases. Most significantly, the ability to simultaneously optimize performance and robustness opens new design paradigms where reliability is built into the optimization process rather than verified post-hoc.

Finally, the ability to model a physics function, rather than an objective function, provides multiple benefits which future work could explore.  For example, the trained surrogate could support optimization with new objective functions without the need for additional simulations. The physical surrogate could even support joint optimization of entirely different design tasks simultaneously. For example, diagnostics and machine protection systems also require detailed tracking, and could build off the same set of simulations, and even the same surrogate model, developed during a \mpbax DAMA optimization. Joint optimization across an entire facility design would provide further improvements in efficiency.

\textit{Acknowledgements} This work was supported by the U.S. Department of Energy, under DOE Contract No. DE-AC02-76SF00515 and the Office of Science, Office of Basic Energy Sciences.

\section*{Appendix A: Algorithm details}
\phantomsection\label{appendix:algorithm}

\subsection{Full multipointBAX algorithm description}

Below is a detailed description of the full \mpbax algorithm, also shown schematically in Fig.~\ref{fig:full_bax_schem}. We also show in algorithm form (Algo.~\ref{alg:mpbax}) along with the hyperparameters we used, along with the detailed surrogate model architectures in the following sections.

\begin{enumerate}
\item Simulate an initial dataset of $n_0$ points for both DA and MA. Each data point in the dataset has input $(c_0, c_1, c_2, c_3, p_0, p_1, \epsilon)$, where $c_i$ is the normalized sextuple strength (ring configuration)\footnote{Note that we have 6 sextuple families to tune, but we fixed the horizontal and vertical chromaticities, leaving only 4 degree of freedom.}, $p_i$ is the position of the particles, and $\epsilon$ is one of $n_\epsilon$ random quadrupole error seeds, and the output is the number of survived turns normalized by the total number of simulated turns. The configuration and position of each data point are randomly selected in their own spaces, and initial dataset for DA and MA are generated independently.

\item Train a surrogate model on the initial dataset. DA and MA have independent NN surrogate models. We train the model for $e_0$ epochs with early-stopping.
\item \textbf{Begin loop:} Run NSGA-II on the surrogate model with population $p$ for $g$ generations, then collect all unique configurations during the run.
\item Use the updated surrogate model to evaluate all configurations found on the ``Pareto front region'' (PFR) in previous iterations. The PFR is defined to include points close to the PF. This reduces variation due to the GA not converging in some iterations. 
\item Select the top $m$ configurations near the PFR found on all previous iterations.

\item Select the set of points that lie near the stability map boundary for each of the selected configurations. The boundary is defined as points where the predicted number of turns lies between $t_\mathrm{min}$ and $t_\mathrm{max}$, normalized by the max number of turns simulated.
The ranges are chosen to balance the number of in-boundary and out-boundary points\footnote{We use cutoff values in DA/MA area calculations of 0.75 for DA and 0.94 for MA. A strict 1.0 cutoff (as used in a physics simulation) cannot be used for the surrogate model both because of noisy predictions and because the sigmoid output of the model never reaches exactly 1. The chosen thresholds were set to roughly align the evaluated objectives with those from the NN predictions.}.
\item Select $n$ points from the set to simulate.
\item Retrain the surrogate models for $e$ epochs.
\item Return to step 3, unless $L$ loops have been completed.
\item Select $m_f$ final configurations predicted to lie on the Pareto front (for HV calculation) or points predicted to maximize a linear combination of DA and MA.
\item Evaluate the selected DA and MA maps with simulation on a new set of $n_\epsilon$ rings (different error seeds from the ones used during the BAX run), calculate the DA and MA areas for each ring, averaging over the values to get the mean DA and MA areas, and return as the final metric.

\end{enumerate}

Hyperparameters are given in Table ~\ref{tab:bax_hyperparameters}. These hyperparameters have not been optimized extensively, which could further improve efficiency. 

\begin{table}[h]
\centering
\begin{tabular}{lcc}
\toprule
\textbf{Description} & \textbf{Symbol} & \textbf{Value} \\
\midrule
NSGA-II generations & $g$ & 20 \\
NSGA-II population size & $p$ & 200 \\
Configurations selected each loop & $m$ & 300 \\
Initial points simulated & $n_0$ & 3000 \\
Points simulated each loop & $n$ & 50 \\
Boundary threshold minimum (DA/MA) & $t_\mathrm{min}$ & 0.4/0.85 \\
Boundary threshold maximum (DA/MA) & $t_\mathrm{max}$ & 0.75/0.95 \\
Points in a full DA and MA map & $n_\mathrm{sim}$ & 2397 \\
Initial training epochs & $e_0$ & 150 \\
Training epochs each loop & $e$ & 10 \\
Number of loops & $L$ & 2000 \\
Number of error seeds & $n_\epsilon$ & 10 \\
Final configurations evaluated & $m_f$ & 10-20 \\
\bottomrule
\end{tabular}
\caption{\mpbax hyperparameters used in the SSRL-X example}
\label{tab:bax_hyperparameters}
\end{table}

\begin{figure}
\vspace{-4mm}
\begin{algorithm}[H]
\caption{\mpbax algorithm for DAMA optimization}
\label{alg:mpbax}
\begin{algorithmic}[1]

\State{\textcolor{gray}{\texttt{\# $g$:\,generations in NSGA-II, $p$:\,population size in NSGA-II, $t_{min},t_{max}$:\,minimum and maximum turns survived to define boundary region, $m$:\,\#\,\,of configurations selected each iteration, $n$:\,\#\,\,of particles to simulate in each iteration, $e$:\,epochs to train each iteration, $L$:\,\#\,\,of iterations. Subscripts $0$ and $f$ for $n$, $m$, and $e$, denote the value for the initialization and final evaluation steps.}}}
\State \textbf{Input:} $g, p, t_{\min}, t_{\max}, m, m_0, m_f, n, n_0, e, e_0, L$
\State{\textcolor{gray}{\texttt{\# Initialize random configuration set:  }}}
\State $\mathcal{C}_0 = \{c_1, \dots, c_{m_0}\}$
\State{\textcolor{gray}{\texttt{\# Choose random points from initial configurations: }}} 
\State $\mathcal{S}_0 = \{x_c, z_c \mid x_c \sim \text{Uniform}(\mathcal{X}_c), z_c \sim \text{Uniform}(\mathcal{Z}_c),\ \forall\, c \in \mathcal{C}_0\}$ with $|\mathcal{S}_0| = n_0$
\State{\textcolor{gray}{\texttt{\# Initialize training dataset by simulating sampled points:}}}
\State $\mathcal{D} \gets \{(x, z, f_{\text{DA}}(x,c), f_{\text{MA}}(z,c)) \mid x, z \in \mathcal{S}_0\}$
\State{\textcolor{gray}{\texttt{\# Train NN surrogates $\hat{f}_{\text{DA}}, \hat{f}_{\text{MA}}$ using $\mathcal{D}$ for $e_0$ epochs}}}
\State $\hat{f}_{\text{DA},0}, \hat{f}_{\text{MA},0} \gets \text{Train}( \mathcal{D}_0, e_0)$

\For{$\ell = 1$ to $L$}
    \State{\textcolor{gray}{\texttt{\# Find Pareto front by running NSGA-II on NN surrogates}}}
    \State $\mathcal{P}_\ell \gets \text{NSGA-II}(\hat{f}_{\text{DA},\ell-1}, \hat{f}_{\text{MA},\ell-1}, g, p)$
    \State{\textcolor{gray}{\texttt{\# Select up to $m$ Pareto-optimal configurations from all previous loops:}}}
    \State  $\mathcal{C}_\ell \subset \mathcal{P}_\ell$ with $|\mathcal{C}_\ell| \leq m$
    \State{\textcolor{gray}{\texttt{\# Identify set of boundary points for each configuration:}}}
    \State $\mathcal{B}_{x,c} \gets \{x \mid t_{\min} \leq \hat{f}_{\text{DA},\ell-1}(x,c) \leq  t_{\max} \}, \, \mathcal{B}_{z,c} \gets \{z \mid t_{\min} \leq \hat{f}_{\text{MA},\ell-1}(z,c) \leq  t_{\max} \}, \, \forall \, c \in  \mathcal{C}_\ell$ 
    \State{\textcolor{gray}{\texttt{\# Select points at random from set of boundary points}}}
    \State $\mathcal{S}_\ell \subset \text{Uniform}(\mathcal{B}_x, \mathcal{B}_z)$ with $|\mathcal{S}_\ell| = n$
    \State{\textcolor{gray}{\texttt{\# Run simulation on selected points and update dataset}}}
    \State $\mathcal{D}_\ell \gets \mathcal{D}_{\ell-1} \cup \{(x, z, f_{\text{DA},\ell-1}(x,c), f_{\text{MA},\ell-1}(z,c)) \mid x, z \in \mathcal{S}_\ell\}$
    \State{\textcolor{gray}{\texttt{\# Retrain NN on $\mathcal{D}$ for $e$ epochs}}}
    \State $\hat{f}_{\text{DA},\ell}, \hat{f}_{\text{MA},\ell} \gets \text{Train}(\hat{f}_{\text{DA},\ell-1}, \hat{f}_{\text{MA},\ell-1}, \mathcal{D}_\ell, e)$
\EndFor
\State{\textcolor{gray}{\texttt{\# Find Pareto front running NSGA-II on final NN surrogate}}}
\State $\mathcal{P}_f \gets \text{NSGA-II}(\hat{f}_{\text{DA,L}}, \hat{f}_{\text{MA,L}}, g, p)$
\State{\textcolor{gray}{\texttt{\#  Select $m_f$ configurations from final surrogate Pareto front: }}}
\State $\mathcal{C}_f \subset \mathcal{P}_f$ with $|\mathcal{C}_f| \leq m_f$
\State{\textcolor{gray}{\texttt{\#  Select all points from the chosen configurations: }}}
\State $\mathcal{S}_f = \{x_c, z_c \in \mathcal{X}_c,\mathcal{Z}_c, \,\forall\, c \in \mathcal{C}_f\}$
\State{\textcolor{gray}{\texttt{\#  Simulate true objectives: }}}
\State $\mathcal{D}_f = \{(f_{\text{DA}}(x,c), f_{\text{MA}}(z,c)) \forall\, x,z \in \mathcal{S}_L\}$
\State{\textcolor{gray}{\texttt{\# Compute and return metric, e.g., hypervolume:}}}
\State $HV \gets \text{Hypervolume}(\mathcal{D}_f)$

\end{algorithmic}
\end{algorithm}
\end{figure}

\subsection{Acquisition function}
\label{appendix:acquisition-function}

BAX is a general framework for algorithm-driven acquisition, and there are various strategies to implement the acquisition function. The information-based `InfoBAX’ has good theoretical motivation, but is onerous computationally \cite{neiswanger2021bayesian}. Each InfoBAX iteration requires: taking function samples from the model posterior; adding $n$ points from running the algorithm on these samples to make $n$ new datasets; and finally retraining the probabilistic model for all $n$ datasets. 
As a result, NN-based InfoBAX requires training an ensemble of probabilistic NNs at each iteration, and the training cost could become significant. 

An alternative acquisition function `MeanBAX’ has less theoretical basis, but is simpler and more computationally efficient \cite{chitturi2024targeted}. MeanBAX is also more exploitative than InfoBAX, targeting observations directly towards the best estimate of the user objective. In MeanBAX, the BAX algorithm is run on the model's posterior mean prediction, which identifies configurations that satisfy the user's specified criteria. In the nominal MeanBAX approach, the point(s) in this selected set with the highest predicted uncertainty (posterior variance) are chosen, requiring, e.g. a single neural ensemble. In the DAMA case, we simplify even further,
using a single, deterministic network. We implement MeanBAX by first selecting configurations that are predicted to lie on or near the Pareto front (running the algorithm on our model), and then selecting individual points that are predicted to lie near the DA and MA boundaries for the chosen configurations. 

Due to availability of CPU clusters, as well as overhead of NN training and GA optimization, it is preferable from a wall-clock perspective to simulate a batch of physics simulations in parallel. While there are established methods of batch selection in a BO context (e.g. hallucination~\cite{kriging-believer}) these methods are not intended for large batches of dozens of points. Instead, we employ a simple heuristic: the MeanBAX criteria described above result in the selection of approximately 2500 points across 250-300 configurations, from which we randomly select a batch of 50 points to simulate.  A more principled approach to selecting the batch, e.g. through joint mutual information, would further improve efficiency. 

\subsection{Neural network architecture}
\label{appendix:nn-structure}

\begin{figure}[htbp]
    \centering
    \includegraphics[width=\linewidth]{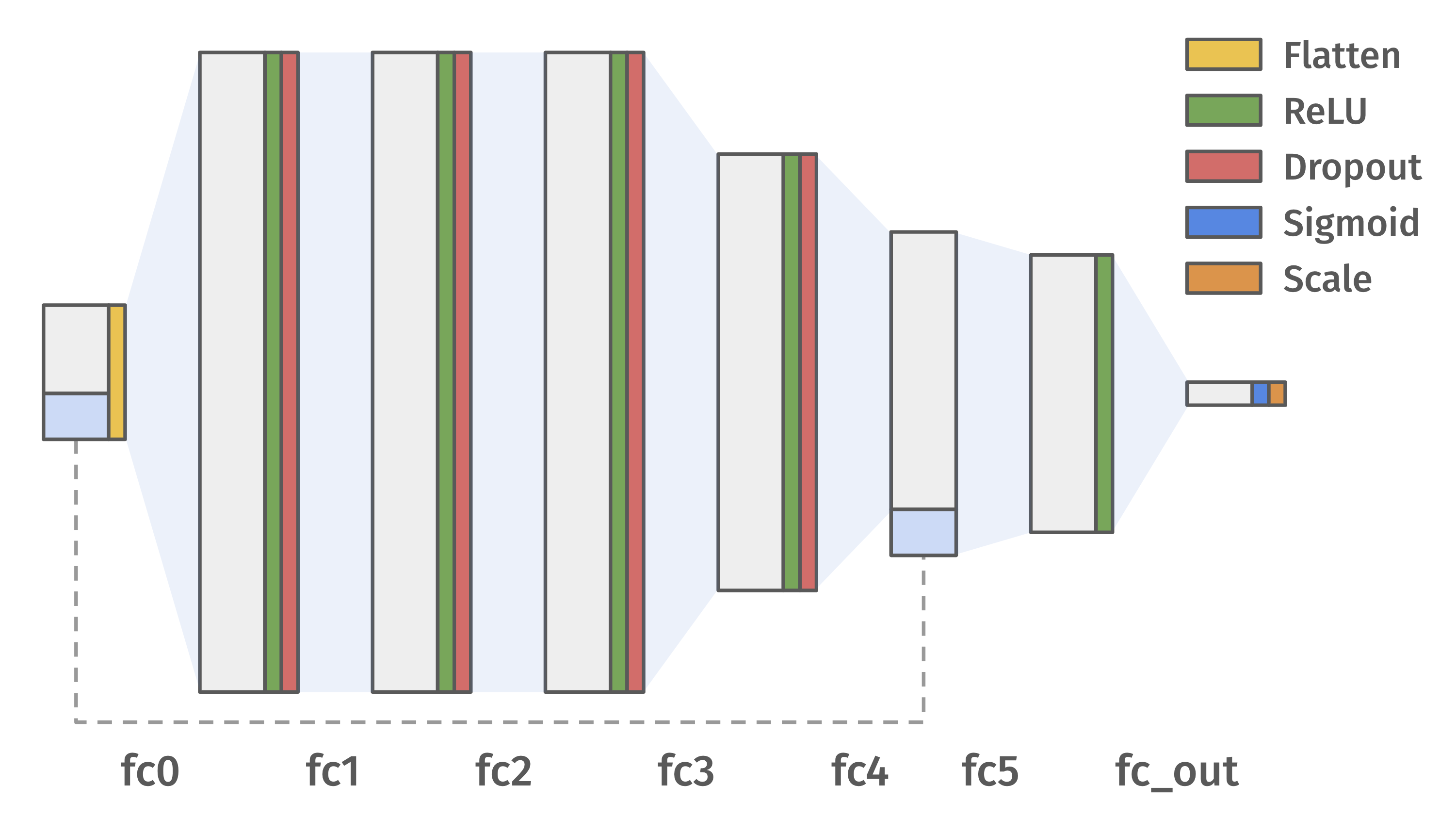}
    \caption{Neural net architectures for DA and MA. The fc layers denoted by shaded blue area are fully connected layers. The lightgray blocks show the change of the input data shape when it goes through the layers. The lightblue block is the particle position part of the input. We concatenate the input again with the fc4 layer output to weight the inputs more, although the observed impact on training was not significant.}
    \label{fig:NN_architecture}
\end{figure}

For the NN we use a neural field architecture, shown in figure \ref{fig:NN_architecture} with values given in Table \ref{tab:NN_hyperparams}.

\begin{table}[htbp]
    \centering
    \begin{tabular}{cccc}
        \toprule
        \textbf{Layer} & \textbf{Input Size} & \textbf{Output Size} & \textbf{Operation} \\
        \midrule
        Flatten        & 6                   & 6                    & - \\
        fc0            & 6                   & 800                  & ReLU \\
        fc1            & 800                 & 800                  & ReLU \\
        fc2            & 800                 & 800                  & ReLU \\
        fc3            & 800                 & 400                  & ReLU \\
        fc4            & 400                 & 200                  & ReLU \\
        fc5            & 202                 & 200                  & ReLU \\
        fc\_out        & 200                 & 1                    & Sigmoid, scaled by $1.0$ \\
        Dropout        & -                 & -                  & $p = 0$ \\
        Concat         & 200 + 2             & 202                  & Concatenation \\
        \bottomrule
        \end{tabular}
    \caption{NN model parameters used for both models.}
    \label{tab:NN_hyperparams}
\end{table}

We control the dropout rate in the dropout layers in the pre-training and trainings during BAX loops to mitigate overfitting and underfitting. The last scale operation ensures the model has the capability to output the whole range of the normalized survived turns.

\section*{Appendix B: Performance comparison on an alternative metric}
\phantomsection\label{appendix:single-metric}

As an additional comparison, we evaluate a single-configuration metric defined as a linear combination of the DA and MA metrics,
\[
M = M_\text{DA} + \alpha M_\text{MA}.
\]
This scalar indicator broadly mirrors HV while reducing final-evaluation cost, since only one configuration needs to be tested rather than the full Pareto front across ten error seeds. We set $\alpha=100$ to reflect the relative magnitudes of the optimal DA and MA contributions, and we also report $\alpha=50$ and $\alpha=200$ (half and double) to weigh either DA or MA more heavily. Because the MA contribution is relatively flat along the Pareto front, these single-configuration selections naturally lean toward the DA-optimal side.  

\begin{figure}[htbp]
    \centering
    \includegraphics[width=\linewidth]{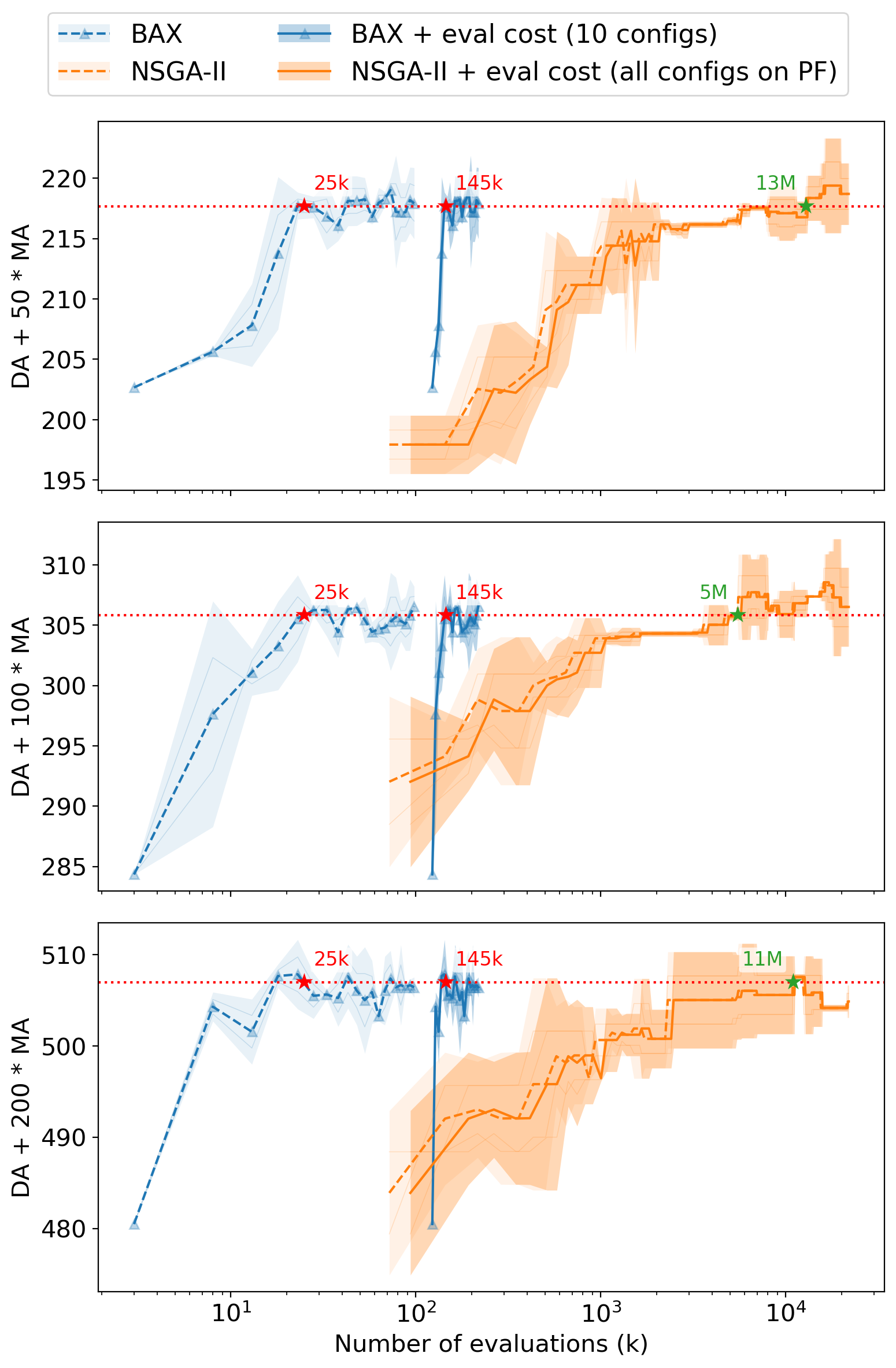}
    \caption{BAX vs. NSGA-II performance as a function of iteration number using the single-configuration metric $M = M_\text{DA} + \alpha M_\text{MA}$ with $\alpha=50$ (top), 100 (middle), and 200 (bottom). Mean performance over two runs (dashed and solid curves) with $2\sigma$ error bands (shaded areas) is shown. Dashed lines denote optimization cost, solid lines include the cost of the final evaluation. Red stars indicate \mpbax\ early-stopping points, and green stars show where NSGA-II matches that performance.}
    \label{fig:BAX_vs_SoTA_LCO}
\end{figure}

For cost accounting, Figure~\ref{fig:BAX_vs_SoTA_LCO} reports where NSGA-II first matches the \mpbax\ convergence level: roughly $13\text{M}$, $5\text{M}$, and $11\text{M}$ evaluations for $\alpha=50,100,200$, respectively. For \mpbax, we include only a modest final verification step by evaluating $10$ candidate configurations (adding $\sim\!145\text{k}$ evaluations), which halves the cost compared to evaluating the entire Pareto front. In contrast, for NSGA-II we evaluate all Pareto-front configurations (typically $\sim\!30$), as this additional cost is negligible compared to its multi-million-evaluation runs. This accounting slightly disadvantages \mpbax, yet the conclusion remains robust: across $\alpha$, \mpbax\ achieves comparable performance with up to nearly 100x reductions in evaluation cost.

\section*{Appendix C: Early stopping indicator for BAX}
\phantomsection\label{appendix:early-stopping}

\begin{figure}[htbp]
    \centering
    \includegraphics[width=\linewidth]{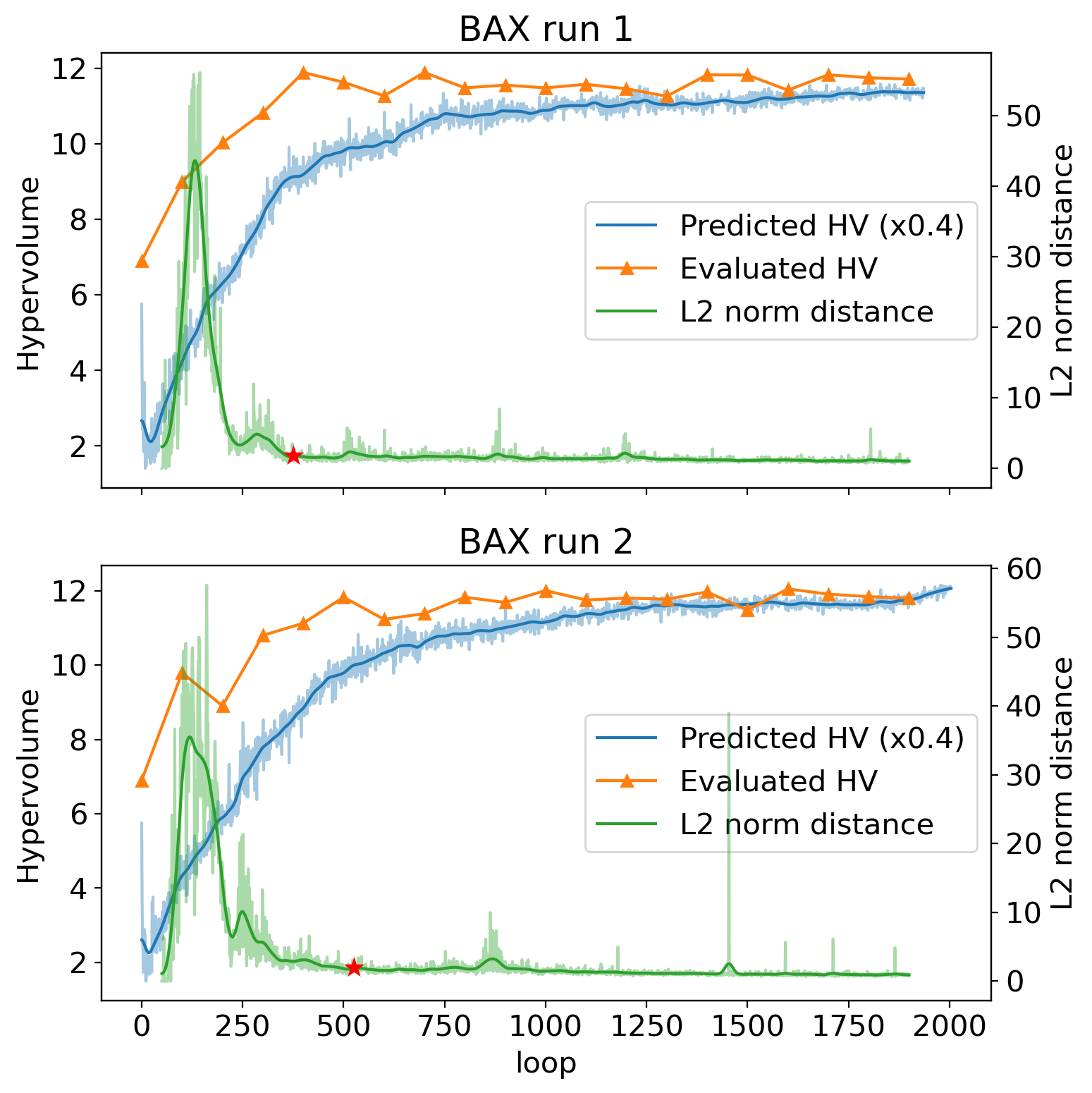}
    \caption{Behavior of the early-stopping indicator for two BAX runs. The orange curve shows the evaluated HV (used as the ground truth for convergence but unavailable during the run), and the blue curve shows the predicted HV of the selected configurations, which continues rising and provides no clear turning point. The green curve is the L2 distance in predicted metrics comparing the current run to the prediction from 50 loops prior. As the BAX model stops learning, this distance decreases. We find the convergence behavior aligns well with the ground truth convergence.}
    \label{fig:early_stopping}
\end{figure}

The early-stopping criterion is based on the $L_2$ norm of the difference between the predictions of the DA and MA objectives on the configurations in the most recent predicted Pareto-front region, evaluated by the current loop's models and earlier models, e.g. from 50 loops prior (Fig.~\ref{fig:early_stopping}). Specifically, for each configuration in the predicted Pareto-front region, both sets of models produce a 2D prediction \((M_\text{DA}, M_\text{MA})\); we compute the $L_2$ distance between the two predictions for each configuration and take the mean distance as the indicator value. This measures how much the predicted Pareto-front region shifts between iterations and serves as a proxy for model convergence.

Once the smoothed indicator falls below a set threshold (here 1.7 chosen empirically) and remains stable, the corresponding loop is taken as the termination point. Additional loops are carried out beyond this point to confirm the validity of the indicator.

Note that the evaluated HV is systematically lower than the predicted HV in Fig.~\ref{fig:early_stopping}. This arises from two factors: (i) different thresholds are used for DA and MA calculations in the BAX run (with a lower cutoff to account for model limitations) and in the evaluation (where a strict threshold of 1.0 is applied), and (ii) the evaluation HV is obtained by selecting only the 20 best predicted configurations, while the predicted HV is computed from the entire predicted Pareto front. To place the two on a comparable scale in the same plot, we apply a scaling factor of 0.4 to the predicted HV values.

\section*{Appendix D: Average time cost per tracking for BAX and NSGA-II runs}
\phantomsection\label{appendix:time-cost}

Each particle-tracking run varies in duration, as simulations of particles that survive more turns take longer to complete. Because BAX tends to query points near or inside the survival boundary, where longer lifetimes are common, its average cost per simulation may be higher than that of NSGA-II. We benchmarked the end-to-end runtime on a 128-core cluster node and examined its dependence on the number of survived turns. Although runs with the same number of turns show a long-tailed spread in runtime—likely due to core and thread orchestration—the median runtime reveals a clear linear trend, approximately following $t \approx 3\cdot\hat{n}+0.5$, where $t$ is the time cost in second, and $\hat{n}$ is the normalized number of survived turns.

\begin{figure}[htbp]
    \centering
    \includegraphics[width=\linewidth]{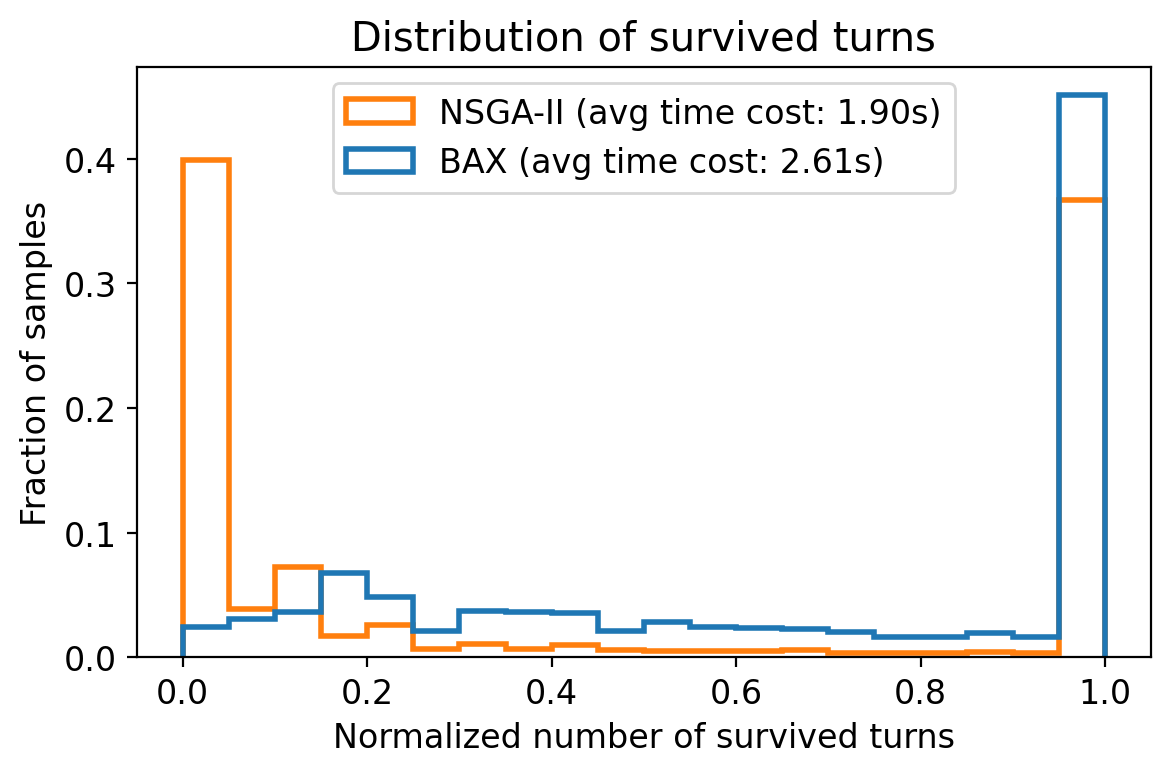}
    \caption{Distribution of the normalized number of survived turns for BAX and NSGA-II runs. BAX has higher density with an intermediate number of survived turns, reflecting its focus on boundary points
    whereas NSGA-II evaluates all sampling positions across the DA/MA map.}
    \label{fig:time_cost_per_tracking}
\end{figure}

Using this relation, we estimated the average runtime per simulation from all BAX and NSGA-II runs, as shown in Fig.~\ref{fig:time_cost_per_tracking}. The resulting averages are \SI{2.61}{\second} for BAX and \SI{1.90}{\second} for NSGA-II, corresponding to a factor of 1.37 difference. While this overhead is non-negligible, it does not materially affect the conclusions of this work.

\FloatBarrier
\bibliographystyle{apsrev4-2}
\bibliography{accel}

\end{document}